\documentclass[aps,prx,superscriptaddress,twocolumn,twoside,nofootinbib,floatfix,a4paper]{revtex4-2}
\pdfoutput=1

\usepackage{graphicx}
\usepackage{dcolumn}
\usepackage{bm}
\usepackage{dsfont}
\usepackage[utf8]{inputenc}
\usepackage{amsfonts}
\usepackage[normalem]{ulem}
\usepackage{comment}
\usepackage{xcolor}
\usepackage{amsthm}
\usepackage{mathtools}
\usepackage{bbm}
\usepackage{bm}
\usepackage{graphicx}
\usepackage{tikz}
\usetikzlibrary{arrows}
\usepackage{soul}
\usepackage{multirow}
\usepackage{physics}
\usepackage{makecell}
\usepackage{dsfont}
\usepackage{placeins}

\usepackage{colortbl}
\definecolor{chshcol}{RGB}{180,220,255}

\usepackage[colorlinks=true,linkcolor=blue,citecolor=blue,urlcolor=blue]{hyperref}

\begin{document}

\title{Almost all pure entangled states enable unbounded nonlocality sharing}


\author{Anna Steffinlongo}
    \affiliation{ICFO - Institut de Ciencies Fotoniques, The Barcelona Institute of Science and Technology, 08860 Castelldefels, Spain}

\author{Nicola D’Alessandro}
    \affiliation{Physics Department and NanoLund, Lund University, Box 118, 22100 Lund, Sweden}

\author{Martin J. Renner}
    \affiliation{ICFO - Institut de Ciencies Fotoniques, The Barcelona Institute of Science and Technology, 08860 Castelldefels, Spain}

\date{\today}

\begin{abstract}
We establish a connection between Hardy's paradox and nonlocality sharing in sequential bipartite scenarios, where each subsystem is measured in turn by a chain of observers. We show that any correlations exhibiting a Hardy paradox in the two-input two-output scenario enable sequential violations of the CHSH inequality between arbitrarily many pairs of observers, using only projective measurements and the assistance of a small local ancilla system. Since almost all pure entangled states, with the only exception of the maximally entangled one, admit Hardy correlations, our protocol applies generically: almost all pure entangled states, if assisted by a local ancilla, allow for nonlocality sharing between arbitrarily many observer pairs using only projective measurements.

\end{abstract}

\maketitle

\section{Introduction}

Bell nonlocality is a defining feature of quantum theory, revealing correlations between spatially separated systems that cannot be explained by any local hidden-variable model~\cite{Bell, Brunner_2014}. Beyond its foundational significance, Bell nonlocality constitutes a key resource for device-independent quantum information processing, enabling tasks such as secure communication~\cite{Ekert91, Acin07} and certified randomness generation~\cite{pironio10}.

Standard Bell scenarios consider a fixed set of parties, each performing one measurement per experimental round on their share of an entangled state. In this framework, measurements are effectively destructive: once a system is measured, it is no longer available to generate further nonlocal correlations.
Recent work has begun to challenge this perspective by considering sequential measurement scenarios, in which quantum systems are passed from one observer to another. Surprisingly, in such settings multiple independent observers can violate a Bell inequality using the same physical systems, a phenomenon known as nonlocality sharing~\cite{Cai2025Review}. Silva et al.\ showed that two sequential Bobs can share nonlocality with a single Alice~\cite{Silva2015}, while Brown and Colbeck demonstrated that an arbitrary number of one-sided observers can do so~\cite{Brown2020}. These findings have been extended to multipartite settings~\cite{Saha2019QuIP, Zhang2021PRA, Xi2023PRA, Shen2024PRA}, network scenarios~\cite{Mahato2022PRA, Mao2023PRR}, and realized in experiments~\cite{Schiavon2017experiment, Hu2016experiment, Feng2020experiment, Zhu2022experiment,  Xiao2024experiment}.
A particularly striking aspect of nonlocality sharing is that it can be realized using only projective measurements~\cite{Steffinlongo2022,Sasmal2024}. Indeed, projective measurements are typically regarded as maximally disturbing, in contrast to the weak or unsharp measurements, which are specifically engineered to preserve coherence for subsequent observers.

Despite the progress, existing protocols are typically limited in at least one aspect~\cite{Silva2015, Brown2020, Cabello2021, Steffinlongo2022,Sasmal2024,ZhangPRA2024}: they either consider only one-sided sequential scenarios, rely on specific quantum states, or employ unsharp measurements. Together with the strong evidence that two-sided sharing with projective measurements is impossible for qubit systems~\cite{Cheng2021,Cheng2022}, those results leave open the question of whether two-sided nonlocality sharing with projective measurements is possible in general.


In this Letter, we close this gap by establishing a general connection between nonlocality sharing and Hardy's paradox~\cite{Hardy1992, hardy1993, Goldstein1994, Rabelo2012, Jiang2018PRL}. Hardy's paradox provides a particularly sharp manifestation of quantum nonlocality, revealing a contradiction with local realism without invoking Bell inequalities. We show that any correlations exhibiting a Hardy paradox (quantum or otherwise) can be used to build a sequential scenario in which arbitrarily many distant observer pairs violate the CHSH inequality~\cite{CHSH}. Our protocol is notably simple: it requires only two-qubit states assisted by one local ancilla on each side, the simplest Bell inequality, standard projective measurements, and produces violations for every Alice--Bob pair simultaneously. Since Hardy's paradox holds for all pure non-maximally entangled states~\cite{Hardy1992, hardy1993, Chen2013}, our result implies that two-sided nonlocality sharing with projective measurements is a generic feature of quantum entanglement rather than an exception. Our findings extend even beyond quantum theory to any generalized probabilistic theory~\cite{Muller2020, Plavala2023} admitting a Hardy's paradox~\cite{popescu_quantum_1994, Kolangatt2025}. 

\section{Preliminaries}
\subsection{Nonlocality Sharing Scenario}\label{sec:scenario}

\begin{figure*}[hbt!]
    \centering
    \includegraphics[width=1.0\linewidth]{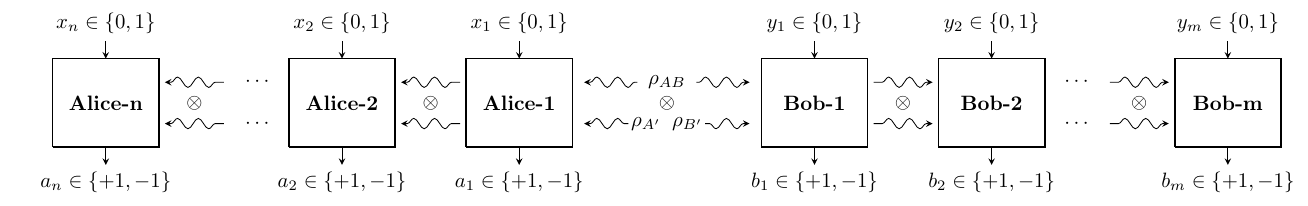}
    \caption{Sequential scenario considered in this work. A source distributes a pure entangled state together with two quantum ancillas $\rho = \rho_{AB}\otimes\rho_{A'}\otimes\rho_{B'} = \ketbra{\Psi}{\Psi}_{AB}\otimes\ketbra{no}{no}_{A'}\otimes\ketbra{no}{no}_{B'}$ to chains of observers on Alice's and Bob's sides. For convenience, the ancillas are included in the initial state distributed by the source, so that all observers perform the same operation. Equivalently, they could be prepared locally by the first observers. Each observer may perform a projective measurement or pass the system to the next observer.}
    \label{setup}
\end{figure*}

We consider a bipartite sequential scenario in which a central source distributes a pure entangled state $\ket{\Psi}_{AB}\in \mathbb{C}^d\otimes \mathbb{C}^d$ together with two local ancillas, one for each branch. The ancillas are initialized in a fixed reference state $\ket{no}$, so that the global initial state is $\rho = \rho_{AB}\otimes\rho_{A'}\otimes\rho_{B'} = \ketbra{\Psi}{\Psi}_{AB}\otimes\ketbra{no}{no}_{A'}\otimes\ketbra{no}{no}_{B'}$.
Throughout the protocol, each ancilla records the local measurement history of its corresponding subsystem.
We label the observers of the two branches Alice-$i$ and Bob-$i$, ordered by index $i$ as illustrated in Fig.~\ref{setup}.

Each Alice-$i$ (Bob-$i$) receives the local system together with its ancilla and a randomly drawn input $x_i\in\{0,1\}$ ($y_i\in\{0,1\}$). The observer performs a local operation on the joint system and ancilla, obtaining an outcome $a_i\in\{+1,-1\}$ ($b_i\in\{+1,-1\}$). In our implementation, this operation consists of accessing the ancillary register via a measurement in the computational basis and subsequently performing a corresponding projective measurement on the system $\rho_{AB}$. 
After the measurement, the ancilla is locally updated and the resulting joint state is forwarded to the next observer.
This results in correlations taking the form $p(a_1,\dots,a_n; b_1,\dots,b_m \mid x_1,\dots,x_n; y_1,\dots,y_m)$.

For every pair of observers (Alice-$i$, Bob-$j$) we ask whether the two-party marginal distribution $p(a_i,b_j|x_i,y_j)$, obtained by marginalizing over the outcomes of all other observers and averaging over their inputs, admits the local hidden variable (LHV) decomposition
\begin{align}\label{eq:local}
p(a_i,b_j|x_i,y_j)=
\sum_{\lambda} p(\lambda)\,
p_A(a_i|x_i,\lambda)\,
p_B(b_j|y_j,\lambda).
\end{align}
In order to show that they do not, and hence that nonlocality sharing is possible between any pair of observers, each pair tests the Clauser--Horne--Shimony--Holt (CHSH) inequality~\cite{CHSH}:
\begin{align}\notag
    S_{i,j}=\frac{1}{4}\Big(&p(a_i\neq b_j|00)+p(a_i=b_j|01)\\&+p(a_i=b_j|10)+p(a_i=b_j|11)\Big)\leq \frac{3}{4}\label{eq:chsh}
\end{align}
which satisfies the local bound $S \leq 3/4$~\cite{Brunner_2014}. 

\subsection{Hardy's paradox}\label{sec:preliminaries}

One of the simplest demonstrations of nonlocality and central to this work is Hardy's paradox. Consider correlations satisfying the following conditions:
\begin{align}
    p(a=-1,b=-1|x=0,y=0)&=0,\label{eq:hardy1}\\
    p(a=-1,b=+1|x=1,y=0)&=0,\label{eq:hardy2}\\
    p(a=+1,b=-1|x=0,y=1)&=0,\label{eq:hardy3}\\
    p(a=-1,b=-1|x=1,y=1)&\eqqcolon\gamma>0 \, .\label{eq:hardy4}
\end{align}
The first three conditions impose constraints that are incompatible with the last one under any LHV model; a proof is given in Appendix~\ref{App:Hardyproof} for completeness.\footnote{Alternatively, the nonlocality of Hardy correlations given in Tab.~\ref{tab:hardy} can also be certified via the CHSH inequality in Eq.~\eqref{eq:chsh}, which gives a value of $3/4+\gamma/2>3/4$.}

Hardy-type correlations appear in quantum theory, but our arguments do not only apply to quantum correlations but to general nonsignalling correlations, i.e., correlations satisfying
\begin{align}
\sum_b p(a,b|x,y)&=\sum_b p(a,b|x,y') \quad \forall\, a,x,y,y',\\
\sum_a p(a,b|x,y)&=\sum_a p(a,b|x',y) \quad \forall\, b,x,x',y, \label{eq:nonsignal}
\end{align}
so that one party's marginal distribution does not depend on the other party's input. Under these conditions given by Eq.~\eqref{eq:hardy1}-\eqref{eq:nonsignal}, in the two-input two-output scenario Hardy-like correlations are parametrized by five degrees of freedom, as shown in Table~\ref{tab:hardy}. A Hardy paradox occurs whenever $\gamma > 0$.

\begin{table}[h!]
\centering
\small
\begin{tabular}{cc|cc|cc}
 & & \multicolumn{2}{c|}{$y=0$} & \multicolumn{2}{c}{$y=1$} \\
 & & $b=+1$ & $b=-1$ & $b=+1$ & $b=-1$ \\ \hline
\multirow{2}{*}{$x=0$}
 & $a=+1$ &  $1-\zeta$ & $\beta+\gamma+\varepsilon$ & $1-\alpha-\gamma-\delta$ & $0$ \\
 & $a=-1$ & $\alpha+\gamma+\delta$ & $0$ & $\delta$ & $\alpha+\gamma$ \\ \hline
\multirow{2}{*}{$x=1$}
 & $a=+1$ & $1-\beta-\gamma-\varepsilon$ & $\varepsilon$ & $1-\alpha-\beta-\gamma$ & $\alpha$ \\
 & $a=-1$ & $0$ & $\beta+\gamma$ & $\beta$ & $\gamma$ \\
\end{tabular}
\caption{Generic Hardy-type probability table for binary inputs $x,y$ and binary outputs $a,b$, with parameters $\alpha,\beta,\gamma,\delta,\varepsilon\in[0,1]$ chosen such that all entries are valid probabilities. We set $\zeta:=\alpha+\beta+2\gamma+\delta+\varepsilon$ for convenience. The remaining entries follow from normalization and nonsignalling. A Hardy paradox occurs whenever $\gamma>0$.}
\label{tab:hardy}
\end{table}

Hardy's paradox occurs for every pure non-maximally entangled two-qubit state and can be generalized to arbitrary dimensions~\cite{Goldstein1994, Chen2013}. Appendix~\ref{App:Hardyforquantum} provides an explicit construction of Hardy correlations for any non-maximally entangled pure state of arbitrary dimension.

To give an example, consider the two-qubit case. Up to local unitaries, a partially entangled two-qubit state can be written as
\begin{equation}\label{eq:twoqubitstate}
\ket{\Psi}= \cos\theta \left(\frac{\ket{01}+ \ket{10}}{\sqrt{2}}\right) + \sin\theta \ket{00},
\end{equation}
where $\theta \in (0, \pi/2)$ parametrizes the entanglement. The corresponding projective measurements are represented by the sets of projectors $\{M_{a|x}\}_a$ ($\{M_{b|y}\}_b$), such that $M_{a|x}M_{a'|x}=\delta_{aa'}M_{a|x}$ and $
\sum_a M_{a|x}=\mathds{1}$ (and analogously for Bob).
For the state in \eqref{eq:twoqubitstate}, these measurements are (the same for Alice and Bob)
\begin{align}\label{eq:twoqubitmeas}\notag
    M_{+1|0} &= \tfrac{1}{2}\left( \mathbb{I} + \sigma_z \right)=\ketbra{0},\\ 
    M_{+1|1} &= \frac{1}{N}
    \left(\sin\theta\, \ket{0} + \tfrac{\cos\theta}{\sqrt{2}}\, \ket{1}\right)
    \left(\sin\theta\, \bra{0} + \tfrac{\cos\theta}{\sqrt{2}}\, \bra{1}\right),
\end{align}
where $N=\tfrac{1}{2}\cos^{2}\theta + \sin^{2}\theta$ and $M_{-1|x}=\mathds{1}-M_{+1|x}$. The resulting correlations are given by the Born rule
\begin{align}
    p_Q(a,b|x,y)=\Tr[(M_{a|x}\otimes M_{b|y}) \ketbra{\Psi}] \, .
\end{align}
They have the Hardy form of Table~\ref{tab:hardy} with:
\begin{align}
    \alpha = \beta &= 
    \frac{2\cos^{6}\theta}{\big(3-\cos(2\theta)\big)^{2}}\, , 
    &\gamma = 
    \frac{4\sin^{2}\theta\,\cos^{4}\theta}
         {\big(3-\cos(2\theta)\big)^{2}}\, . \label{eq:gamma}
\end{align}
The Hardy probability $\gamma$ vanishes as $\theta\to 0$ or $\theta\to\pi/2$, confirming that the Hardy paradox is absent at those limits~\cite{hardy1993}. For all other values of $\theta$, $\gamma > 0$ and the paradox holds.

\section{Nonlocality sharing from any Hardy correlation}\label{sec:gen}
We now show that all Hardy correlations as in Tab.~\ref{tab:hardy} can be turned into a nonlocality sharing protocol. A more explicit example with two Alices and two Bobs is presented in Appendix~\ref{app:example}. 

For our purposes, it is sufficient to focus on the simplest setting in which all observers on one side have access to the same two dichotomic observables, denoted by $M_0$ and $M_1$. These measurements are precisely those used to realize the Hardy correlations for the given state, e.g. Eqs.~\eqref{eq:twoqubitmeas} for a state in Eq.~\eqref{eq:twoqubitstate}.
The protocol is analogous for both sides, thus we describe it from the perspective of a generic observer, without explicitly distinguishing between Alice and Bob. Since projective measurements typically destroy nonclassical correlations along the sequence, each party may use a bit of local randomness to decide whether to perform the corresponding measurement (and output $a_i$ or $b_i$ according to its outcome) or apply the trivial identity channel $\mathds{1}$, output according to a fixed value, and pass the system unchanged to the next party in the chain. In fact, we distinguish three possible strategies for each observer:
\begin{align}\notag
\text{(I)} \quad & \text{projective measurements are performed for both} \\ \notag
&\text{inputs} \rightarrow \{M_{0}, M_{1}\}, \\\notag
\text{(II)} \quad & \text{input } 0 \text{ is measured, while input } 1 \text{ produces the} \\ \notag
&\text{fixed output } +1 \rightarrow \{M_{0}, \mathds{1}\}, \\\notag
\text{(III)} \quad & \text{input } 0 \text{ produces the fixed output } +1, \text{ while} \\ \label{eq:strategies}
& \text{input } 1 \text{ is measured}\rightarrow \{\mathds{1},M_{1}\}.
\end{align}
To keep track of the strategies used, the local ancilla takes values in the basis $\{\ket{no},\ket0,\ket1\}$, where $\ket{no}$ denotes that no measurement has yet been performed, while $\ket0$ and $\ket1$ indicate that the system has previously been measured in the basis corresponding to input 0 or 1, respectively. The ancilla is initialized in the state $\ket{no}$, and once it leaves this state it remains unchanged.
The role of the ancilla is to prevent the system from being measured twice in incompatible bases. The action performed by observer-$i$ on the state of the ancilla and on the observer's input becomes:
\begin{center}
\begin{tabular}{c|cc}
 & $x,y=0$ & \quad $x,y=1$ \\ \hline 
$\ket{no}$ & $M_0$ & \quad$M_1$ ($p_i$)  or $\mathds{1}$ ($1-p_i$)\\
$\ket0$ & $M_0$ &\quad $\mathds{1}$ \\
$\ket1$ & $\mathds{1}$ &\quad $M_1$
\end{tabular}
\end{center}
where $p_i$ is the probability that the $i$-th observer measures $M_1$ when the ancilla is in the state $\ket{no}$ and the input is $1$.
In this sense, each party performs a mixture of strategy (I) (with probability $p_i$) and (II) (with probability $1-p_i$) if the ancilla is in $\ket{no}$; strategy (II) if $\ket{0}$; and strategy (III) if $\ket{1}$. After applying the corresponding strategy, the observer updates the ancilla whenever a measurement is performed on the system and forwards the joint system to the next observer.


The three strategies induce different effective correlations between a given Alice–Bob pair. We therefore evaluate the resulting correlations for each possible combination of strategies.
Suppose a given pair of Alice and Bob are both allowed to apply strategy (I), i.e. both perform projective measurements for both inputs. Clearly, the resulting correlations for that pair of observers are precisely those given in Tab.~\ref{tab:hardy}. When different strategies are combined, the correlations change. An example of the probability table when strategy~(II) for some Alice-$j$ is combined with strategy~(III) for some Bob-$k$ is shown in Table~\ref{tab:S23}.
\begin{table}[hbt!]
\centering
\begin{tabular}{cc|cc|cc}
 & & \multicolumn{2}{c|}{$y=0$} & \multicolumn{2}{c}{$y=1$} \\
 & & $b=+1$ & $b=-1$ & $b=+1$ & $b=-1$ \\ \hline
\multirow{2}{*}{$x=0$}
 & $a=+1$ & $1-\alpha-\gamma-\delta$ & \cellcolor{chshcol}$0$ & \cellcolor{chshcol}$1-\alpha-\gamma-\delta$ & $0$ \\
 & $a=-1$ & \cellcolor{chshcol}$\alpha+\gamma+\delta$ & $0$ & $\delta$ & \cellcolor{chshcol}$\alpha+\gamma$ \\ \hline
\multirow{2}{*}{$x=1$}
 & $a=+1$ & \cellcolor{chshcol}$1$ & $0$ & \cellcolor{chshcol}$1-\alpha-\gamma$ & $\alpha+\gamma$ \\
 & $a=-1$ & $0$ & \cellcolor{chshcol}$0$ & $0$ & \cellcolor{chshcol}$0$ \\
\end{tabular}
\caption{Probability table when Alice uses strategy~(II) and Bob uses strategy~(III). The CHSH value~\eqref{eq:chsh} (sum of the highlighted values) saturates the local bound $S=3/4$.}
\label{tab:S23}
\end{table}

For all other combinations, the full probability tables are given in Appendix~\ref{app:tables}. Here, we summarize the resulting CHSH values for each combination, see Table~\ref{tab:CHSH}. 

\begin{table}[hbt!]
\centering
\begin{tabular}{c||c|c|c}
 & (I) & (II) & (III) \\ \hline \hline 
 (I)  & $3/4+\gamma/2$ & $3/4$ & $3/4-\beta/2$\\\hline
 (II)  & $3/4$ & $3/4$ & $3/4$\\\hline
 (III)  & $3/4-\alpha/2$ & $3/4$ & $3/4-(\alpha+\beta+\gamma)/2$\\
\end{tabular}
\caption{CHSH values for generic Hardy-type correlations and strategies (I), (II), (III). The crucial structural property is that strategy~(II) always saturates the local bound $S=3/4$, regardless of the strategy used on the opposite side and regardless of the specific values of $\alpha,\beta,\gamma,\delta,\varepsilon$.}
\label{tab:CHSH}
\end{table}

Before turning to the formal proof, it is useful to understand the intuition behind it.

\paragraph{Intuition.}
The key feature of the protocol is the structure of Table~\ref{tab:CHSH}: whenever any party uses strategy~(II), the CHSH value is exactly $3/4$, independently of the other party's strategy. This property is special to Hardy correlations and is the reason why the protocol works. Each observer can default to strategy~(II), which never harms other pairs, and only occasionally use strategy~(I), which generates violations. 

Consider for instance Alice-1: She always receives the ancilla in the state $\ket{no}$, hence she applies either strategy (I) or strategy (II). Whenever she uses strategy (II), she always saturates the local bound with any Bob, no matter which strategy he applies. Hence, she can use strategy (I) with an arbitrary low probability to create a violation with a given Bob as long as he is using strategy (I) sufficiently more often than strategy (III).

To see how violations propagate along the chain, consider Alice-1 and Alice-2. When Alice-1 uses strategy~(I), Alice-2 is forced into strategy~(III) in half of the cases (specifically when $x_1=1$, so that the ancilla is in the state $\ket1$). In those cases, the CHSH value falls below the local bound, and Alice-2 must compensate by using strategy~(I) sufficiently often when she receives the ancilla in $\ket{no}$. This requirement propagates along the chain: each subsequent Alice must use strategy~(I) more frequently to compensate for the cases in which earlier observers have forced her into strategy~(III). An analogous propagation occurs on Bob's side, so that all pairs $(j,k)$ can violate the CHSH inequality simultaneously.

\paragraph{Calculation.}
We now make this precise by computing the CHSH value $S_{(j,k)}$ for any pair Alice-$j$ and Bob-$k$. Let $q_k^{no}$ denote the probability that the state reaches party $k$ still unmeasured, and $q_k^{M_0}$ ($q_k^{M_1}$) the probability that it has already been measured in the basis for input $0$ (respectively $1$). By the construction of the protocol, these probabilities satisfy the recursion
\begin{align}\label{eq:prob_def}\notag
    q_k^{no}&=q_{k-1}^{no}\cdot \frac{1}{2} (1-p_{k-1}),\\ \notag
    q_k^{M_0}&=q_{k-1}^{M_0}+q_{k-1}^{no}\cdot \frac{1}{2},\\
    q_k^{M_1}&=q_{k-1}^{M_1}+q_{k-1}^{no}\cdot \frac{1}{2} p_{k-1},
\end{align}
with $q_1^{no}=1$ and $q_1^{M_0}=q_1^{M_1}=0$. To see this, note that the system remains unmeasured if party $k-1$ receives input $x_{k-1}=1$ (probability $1/2$) and does not measure (probability $(1-p_{k-1})$), similarly for the other expressions. Explicit closed forms are given in Appendix~\ref{app:probabilities}. The probability that party $k$ applies each strategy is then
\begin{equation}
    \begin{pmatrix}
        p_k(\text{I}) \\
        p_k(\text{II})\\
        p_k(\text{III})
    \end{pmatrix} =\begin{pmatrix}
        q_k^{no}\cdot p_k\\
        q_k^{no}\cdot (1-p_k)+q_k^{M_0}\\
        q_k^{M_1}
    \end{pmatrix},
\end{equation}
where $p_k(\text{I})+p_k(\text{II})+p_k(\text{III})=1$. Since the same construction is used for both Alice and Bob, these probabilities can be combined with Table~\ref{tab:CHSH} to obtain the CHSH value for each pair
\begin{align}\label{eq:CHSH}\notag
S_{(j,k)}=& \frac{3}{4} + \frac{\gamma}{2} p_j(\text{I}) p_k(\text{I}) - \frac{\alpha}{2}  p_j(\text{III}) p_k(\text{I}) \\
&- \frac{\beta}{2}  p_j(\text{I}) p_k(\text{III})  -\frac{\alpha+\beta+\gamma}{2}  p_j(\text{III}) p_k(\text{III}).
\end{align}

As shown in Appendix~\ref{app:chsh_proof}, $S_{(j,k)}>3/4$ for all pairs $(j,k)$ whenever
\begin{equation}\label{eq:prob_condition}
    q_i^{no} p_i > 3\frac{\alpha+\beta+\gamma}{\gamma}q_i^{M_1} \quad \text{for } i=j,k.
\end{equation}
This exactly captures the intuition that the probability in which a party uses strategy (I), given by $q_i^{no} p_i$, has to compensate for the cases where the same party has to apply strategy (III), given by probability $q_i^{M_1}$. It turns out that this condition is always satisfiable: choosing
\begin{equation}\label{eq:prob}
    p_i = \frac{\mu^i}{\sum_{l=i}^n\mu^l}=\frac{1-\mu}{1-\mu^{n-i+1}}
\end{equation}
for any $\mu>2+\frac{3(\alpha+\beta+\gamma)}{\gamma}$ guarantees it (see Appendix~\ref{app:prob_choice}), completing the proof that all pairs $(j,k)$ violate the CHSH inequality.

For the two-qubit family of Sec.~\ref{sec:preliminaries}, substituting the parameters $\alpha=\beta$ and $\gamma$ from Eqs.~\eqref{eq:gamma} into this expression gives the minimum $\mu$ required for sequential CHSH violations across all Alice--Bob pairs:
\begin{equation}
\mu > 5 + 3\cot^{2}\theta.
\end{equation}
which is finite for all $\theta \in (0,\pi/2)$.

\section{Discussion}\label{sec:discussion}

\paragraph{Reducing the ancilla to a qubit.}
Our construction uses a three-level ancilla. However, if shared randomness is available, the same protocol can be implemented with a two-level ancilla.
The key idea is to use a shared random variable $\lambda\in\{1,\ldots,n\}$ that selects which observer is allowed to use strategy~(I); all others default to strategy~(II) or~(III) based solely on the ancilla state. The 2-level ancilla, together with the value of $\lambda$, encodes the same information as the 3-level ancilla in the original protocol and therefore leads to the same correlations. The formal equivalence is proved in Appendix~\ref{app:bit_equivalence}, where the explicit form of the probability distribution of the shared randomness, $p(\lambda)$, is derived.

\paragraph{Classical implementation.}
Throughout the protocol, the ancillary system remains diagonal in the computational basis and serves solely as a memory register recording the local measurement history. No coherent superpositions between the basis states $\ket{no}$, $\ket0$ and $\ket1$ are ever created or required. Consequently, the ancilla system may be replaced, without affecting the protocol, by a classical memory storing the same information. This observation applies equally to the qutrit ancilla of the original construction and to the qubit ancilla used together with shared randomness.

\paragraph{Necessity of the memory register.}
One may ask whether the ancillary memory introduced in our protocol is really necessary. The results of Cheng \textit{et al.}~\cite{Cheng2021,Cheng2022} provide strong analytical and numerical evidence supporting the conjecture that two-sided CHSH sharing is impossible when considering a single entangled qubit pair and choosing the measurement settings with equal probability. Even after allowing highly biased measurement selection, together with unsharp measurements implemented via Lüders instruments, they demonstrated two-sided sharing only in a minimal scenario involving two Alices and two Bobs. From this perspective, it is remarkable that the presence of a small memory register - be it quantum or classical - associated with each subsystem allows simultaneous CHSH violations for arbitrarily many observers on both sides.

\paragraph{Experimental outlook.}
Our work shows that introducing only a local memory register, which is physically natural and experimentally straightforward to implement, is sufficient to circumvent these limitations. This observation may pave the way to the first experimental demonstration of two-sided nonlocality sharing. To date, sequential nonlocality sharing with observers on \emph{both} sides has only been realized experimentally for EPR steering~\cite{Zhu2022experiment}, not for Bell nonlocality. Our protocol is therefore well suited to current photonic platforms: beyond the partially entangled state, it requires only standard projective measurements and the local memory register introduced above, which may also be implemented classically~\cite{Xiao2024experiment}.

\paragraph{Generalized probabilistic theories.}
It is worth mentioning that our construction also applies to generalized probabilistic theories (GPT)~\cite{Muller2020, Plavala2023} that obey Hardy's paradox. For instance, the correlations obtained from Popescu–Rohrlich boxes~\cite{popescu_quantum_1994}, which maximally violate the CHSH inequality, are of the form given in Tab.~\ref{tab:hardy} with $\gamma=1/2$ and $\alpha=\beta=\delta=\epsilon=0$. More generally, it is known that GPTs based on polygon state spaces can exhibit a Hardy's paradox~\cite{Kolangatt2025}. In the case of GPTs, the ancillary system is most easily taken to be a classical flag, or any state in the theory capable of transmitting one bit or one trit of information.

\section{Conclusion}
We have shown that nonlocality can be shared among arbitrarily many pairs of observers for almost all pure entangled states--specifically, for all pure states except the maximally entangled one. Our construction relies on the connection between nonlocality sharing and Hardy's paradox: any correlations exhibiting a Hardy paradox allow sequential CHSH violations, and every pure non-maximally entangled state admits Hardy correlations. The resulting protocol shows that a simple local memory register suffices to enable two-sided nonlocality sharing using projective measurements.
It would further be interesting to investigate whether similar ideas extend to multipartite quantum states~\cite{Saha2019QuIP, Zhang2021PRA, Xi2023PRA}, to network nonlocality scenarios~\cite{Mahato2022PRA, Mao2023PRR}, or to other Bell inequalities.


\section*{Acknowledgements}
We thank Carles Roch i Carceller, Antonio Acín and Armin Tavakoli for insightful discussions and comments on the manuscript.
A.S. acknowledges financial support through Ayuda PRE2022-101475 financiada por MCIN/AEI/ 10.13039/501100011033 y por el FSE+. M.J.R. acknowledges financial support through the Juan de la Cierva postdoctoral fellowship (La ayuda JDC2024-055405-I, financiada por MICIU/AEI/10.13039/501100011033 y por el FSE+.). A.S. and M.J.R. acknowledge fincancial support by the Government of Spain (Severo Ochoa CEX2019-000910-S and FUNQIP), Fundaci\'o Cellex, Fundaci\'o Mir-Puig, Generalitat de Catalunya (CERCA program) and the European Union (NEQST 101080086). 
N.D. acknowledges financial support by the Knut and Alice Wallenberg Foundation through the Wallenberg Center for Quantum Technology (WACQT) and the Swedish Research Council under Contract No. 202303498.


\bibliography{bib.bib}

\appendix
\onecolumngrid

\section{Hardy's paradox excludes local hidden-variable models}
\label{App:Hardyproof}

In this Appendix we show that correlations satisfying the Hardy conditions cannot be reproduced by a local hidden-variable model. The Hardy conditions are
\begin{align}
    p(a=-1,b=-1|x=0,y=0)&=0, \label{eq:hardy_corr_1}\\
    p(a=-1,b=+1|x=1,y=0)&=0, \label{eq:hardy_corr_2}\\
    p(a=+1,b=-1|x=0,y=1)&=0, \label{eq:hardy_corr_3}\\
    p(a=-1,b=-1|x=1,y=1)&>0 . \label{eq:hardy_corr_4}
\end{align}

Assume these correlations admit an LHV decomposition $p(a,b|x,y)=\sum_{\lambda} p(\lambda)\,p_A(a|x,\lambda)\,p_B(b|y,\lambda)$. Conditions~\eqref{eq:hardy_corr_1}--\eqref{eq:hardy_corr_3} then imply that for every $\lambda$:
\begin{align}
    \label{constraint1} p_A(a=-1|x=0,\lambda)\,p_B(b=-1|y=0,\lambda)&=0, \\
    p_A(a=-1|x=1,\lambda)\,p_B(b=+1|y=0,\lambda)&=0, \\
    p_A(a=+1|x=0,\lambda)\,p_B(b=-1|y=1,\lambda)&=0 .
\end{align}
Condition~\eqref{eq:hardy_corr_4} implies the existence of some $\lambda=\lambda^*$ with
\begin{align}
    p_A(a=-1|x=1,\lambda^*)\,p_B(b=-1|y=1,\lambda^*)>0,
\end{align}
so both factors must be strictly positive. The second and third constraints then require
\begin{align}
    p_A(a=+1|x=0,\lambda^*)=0, \qquad
    p_B(b=+1|y=0,\lambda^*)=0,
\end{align}
and therefore
\begin{align}
    p_A(a=-1|x=0,\lambda^*)=1, \qquad
    p_B(b=-1|y=0,\lambda^*)=1,
\end{align}
which contradicts condition~\eqref{constraint1}. Hence no LHV model can reproduce correlations satisfying Hardy's conditions.

\section{Hardy's paradox for any pure non-maximally entangled state}\label{App:Hardyforquantum}

To see how to build the Hardy's paradox for any non-maximally entangled pure state, we start by writing the state in its Schmidt form as $\ket{\Psi_{AB}}=\sum_i \sqrt{\lambda_i}\ket{ii}$, with $\lambda_i\geq 0$ and $\sum_i \lambda_i=1$. Suppose there exist $i\neq j$ such that $\lambda_i\neq \lambda_j$ and $\lambda_i,\lambda_j \neq 0$; this holds for every pure state except the maximally entangled one. One can construct measurements yielding Hardy correlations in the two-input, two-output scenario.

Define Alice's measurements as:
\begin{align}
    A_{+|0}=&\frac{1}{2} (\ket{i}+\ket{j})(\bra{i}+\bra{j})+\sum_{k\neq i,j} \ketbra{k}\\
    A_{-|0}=&\frac{1}{2} (\ket{i}-\ket{j})(\bra{i}-\bra{j})\\
    A_{+|1}=&\frac{1}{\lambda_i^2 +\lambda_j^2} (\lambda_i\ket{i}-\lambda_j\ket{j})(\lambda_i\bra{i}-\lambda_j\bra{j})+\sum_{k\neq i,j} \ketbra{k}\\
    A_{-|1}=&\frac{1}{\lambda_i^2 +\lambda_j^2} (\lambda_j\ket{i}+\lambda_i\ket{j})(\lambda_j\bra{i}+\lambda_i\bra{j}),
\end{align}
and Bob's measurements as:
\begin{align}
    B_{+|0}=&\frac{1}{\lambda_i +\lambda_j} (\sqrt{\lambda_i}\ket{i}-\sqrt{\lambda_j}\ket{j})(\sqrt{\lambda_i}\bra{i}-\sqrt{\lambda_j}\bra{j})+\sum_{k\neq i,j} \ketbra{k}\\
    B_{-|0}=&\frac{1}{\lambda_i +\lambda_j} (\sqrt{\lambda_j}\ket{i}+\sqrt{\lambda_i}\ket{j})(\sqrt{\lambda_j}\bra{i}+\sqrt{\lambda_i}\bra{j})\\
    B_{+|1}=&\frac{1}{\lambda_i +\lambda_j} (\sqrt{\lambda_i}\ket{i}+\sqrt{\lambda_j}\ket{j})(\sqrt{\lambda_i}\bra{i}+\sqrt{\lambda_j}\bra{j})+\sum_{k\neq i,j} \ketbra{k}\\
    B_{-|1}=&\frac{1}{\lambda_i +\lambda_j} (\sqrt{\lambda_j}\ket{i}-\sqrt{\lambda_i}\ket{j})(\sqrt{\lambda_j}\bra{i}-\sqrt{\lambda_i}\bra{j}).
\end{align}
A direct computation yields:
\begin{align}
    p(a=-1,b=-1|x=0,y=0)&=0,\\
    p(a=-1,b=+1|x=1,y=0)&=0,\\
    p(a=+1,b=-1|x=0,y=1)&=0,\\
    p(a=-1,b=-1|x=1,y=1)&=\frac{\lambda_i\lambda_j(\lambda_i-\lambda_j)^2}{(\lambda_i^2+\lambda_j^2)(\lambda_i+\lambda_j)}>0.
\end{align}
The construction follows a ladder structure. Starting from $\ket{A_{-|0}}=(\ket{i}-\ket{j})/\sqrt{2}$, the post-measurement state on Bob's side is proportional to $\sqrt{\lambda_i}\ket{i}-\sqrt{\lambda_j}\ket{j}$. Choosing $\ket{B_{-|0}}=(\sqrt{\lambda_j}\ket{i}+\sqrt{\lambda_i}\ket{j})/\sqrt{\lambda_i+\lambda_j}$ orthogonal to this state ensures the first Hardy condition. If instead Alice obtains $\ket{A_{+|0}}$, Bob's state is proportional to $\sqrt{\lambda_i}\ket{i}+\sqrt{\lambda_j}\ket{j}+\sum_{k\neq i,j} a_k \ket{k}$; choosing $\ket{B_{-|1}}=(\sqrt{\lambda_j}\ket{i}-\sqrt{\lambda_i}\ket{j})/\sqrt{\lambda_i+\lambda_j}$ ensures the third condition. Finally, $A_{-|1}$ is fixed by imposing the second condition.

\section{A more explicit example}\label{app:example}
We construct an explicit example with two Alices and two Bobs in which all four CHSH functionals $S_{11},S_{12},S_{21},S_{22}$ exceed the local bound. We present two equivalent implementations of the protocol: one using a qubit ancilla together with shared randomness, and one using a qutrit ancilla with local randomness only.

Consider the two-qubit state
\begin{equation}
    \ket{\psi}_{AB}=\tfrac{1}{\sqrt{3}}\big(\ket{00}+\ket{10}+\ket{01}\big),\qquad
    \rho=\ket{\psi}\!\bra{\psi}_{AB}\otimes
    \ketbra{no}{no}_{A'}\otimes
    \ketbra{no}{no}_{B'}.
\end{equation}

\paragraph{Qubit protocol.}
Each side is equipped with a qubit ancilla, initially in the state $\ket{no}$, together with shared randomness. Each pair (Alice-1, Alice-2) and (Bob-1, Bob-2) shares an independent random variable $\lambda_A,\lambda_B\in\{0,1\}$ with $p(\lambda=0)=r$, $p(\lambda=1)=1-r$. Denoting by $A^i_{x_i}$ the observable used by Alice-$i$, the first parties measure
\begin{equation}
\begin{split}
A^1_0&=Z,\quad A^1_1=\begin{cases}
	X,& \lambda_A=0\\
	\openone,& \lambda_A=1
\end{cases}\\
B^1_0&=Z,\quad B^1_1=\begin{cases}
	X,& \lambda_B=0\\
	\openone,& \lambda_B=1.
\end{cases}
\end{split}
\end{equation}
Alice-2's observable depends on the state of the ancilla and on $\lambda_A$:
\begin{center}
\begin{tabular}{c|cc|cc}
 & \multicolumn{2}{c|}{$\lambda_A=0$} & \multicolumn{2}{c}{$\lambda_A=1$} \\
 & $\ket{0}_{A'}$ & $\ket{1}_{A'}$ & $\ket{0}_{A'}$ & $\ket{1}_{A'}$ \\ \hline
$x_2=0$ & $Z$ & $\openone$ & $Z$ & $Z$ \\
$x_2=1$ & $\openone$ & $X$ & $\openone$ & $X$
\end{tabular}
\end{center}
and symmetrically for Bob-2.

\paragraph{Qutrit protocol.} 
Equivalently, each side uses a qutrit ancilla in the basis $\{\ket{no},\ket{0},\ket{1}\}$, where $\ket{no}$ denotes that the subsystem has not yet been measured, while $\ket{0}$ and $\ket{1}$ indicate that it has already been measured in the $Z$ and $X$ bases, respectively. The corresponding strategies are
\begin{equation}
\text{(I)}=\{Z,X\},\qquad
\text{(II)}=\{Z,\openone\},\qquad
\text{(III)}=\{\openone,X\}.
\end{equation}
Alice-1 (and Bob-1) chooses between strategies (I) and (II) with local probability $r$, while Alice-2 (Bob-2) applies the strategy determined by the incoming ancilla state.

\paragraph{CHSH values.} For both implementations, the resulting CHSH values depend only on the strategies used by Alice and Bob and are given by

\begin{equation}
{\renewcommand{\arraystretch}{1.4}
\begin{array}{c|c|c|c}
 & (\text{I}) & (\text{II}) & (\text{III}) \\ \hline 
 (\text{I})   & \frac{19}{24} & \frac34 & \frac{17}{24} \\\hline
 (\text{II})  & \frac34 & \frac34 & \frac34 \\\hline
 (\text{III}) & \frac{17}{24} & \frac34 & \frac58
\end{array}}
\end{equation}


Averaging over the strategies gives
\begin{align}
    S_{11}(r)&=\frac34+\frac{r^2}{24},\\
    S_{12}(r)=S_{21}(r)&=\frac34+\frac{r}{48}-\frac{r^2}{24},\\
    S_{22}(r)&= \frac34+\frac{1-4r}{96}.
\end{align}
All four exceed the local bound for $0<r<\frac14$. Maximizing $\min_i S_{ii}$ we find the optimal probability $r^\star=\tfrac{\sqrt2-1}{2}\approx 0.207$, which gives
\begin{equation}
	S_{11}(r^\star)=S_{22}(r^\star)\approx 0.7518,\quad
	S_{12}(r^\star)=S_{21}(r^\star)\approx 0.7525.
\end{equation}

\section{Probability tables}\label{app:tables}
We report the probability tables obtained for each combination of strategies chosen by Alice and Bob. Each combination is identified by a pair $(j,k)$ with $j,k \in \{\text{I,II,III}\}$. The parameters $\alpha,\beta,\gamma,\delta,\varepsilon\in[0,1]$ are such that all entries are valid probabilities, and $\zeta:=\alpha+\beta+2\gamma+\delta+\varepsilon$. To see how these correlations are obtained, note that measuring the same system in the same basis by a subsequent observer will reveal the same outcome correlations. For instance, if Strategy (I) for Alice is combined with Strategy (II) for Bob, for the cases $x=0$ and $y=0$ as well as $x=1$ and $y=0$, the correlations are the same as in the original Hardy paradox. For $y=1$, in which case Bob outputs via a fixed strategy $b=+1$, the correlations are then uniquely determined by the no-signalling condition~\eqref{eq:nonsignal}.

From each table, the CHSH value can be computed as the sum of the \colorbox{chshcol}{highlighted} entries:
\begin{equation}
S=\dfrac{1}{4}\Big(p(+,-|0,0)+p(-,+|0,0)+p(+,+|1,0)+p(-,-|1,0)+p(+,+|0,1)+p(-,-|0,1)+p(+,+|1,1)+p(-,-|1,1)\Big),
\end{equation}
yielding the values reported in Table~\ref{tab:CHSH}.

\begin{table}[h!]
\centering
\begin{minipage}{0.48\textwidth}
\centering
\begin{tabular}{cc|cc|cc}
 & & \multicolumn{2}{c|}{$y=0$} & \multicolumn{2}{c}{$y=1$} \\
 & & $b=+1$ & $b=-1$ & $b=+1$ & $b=-1$ \\ \hline
\multirow{2}{*}{$x=0$}
 & $a=+1$ & $1-\zeta$ & \cellcolor{chshcol}$\beta+\gamma+\varepsilon$ & \cellcolor{chshcol}$1-\alpha-\gamma-\delta$ & $0$ \\
 & $a=-1$ & \cellcolor{chshcol}$\alpha+\gamma+\delta$ & $0$ & $\delta$ & \cellcolor{chshcol}$\alpha+\gamma$ \\ \hline
\multirow{2}{*}{$x=1$}
 & $a=+1$ & \cellcolor{chshcol}$1-\beta-\gamma-\varepsilon$ & $\varepsilon$ & \cellcolor{chshcol}$1-\alpha-\beta-\gamma$ & $\alpha$ \\
 & $a=-1$ & $0$ & \cellcolor{chshcol}$\beta+\gamma$ & $\beta$ & \cellcolor{chshcol}$\gamma$ \\
\end{tabular}
\caption{Strategy (I,I).}
\end{minipage}
\hfill
\begin{minipage}{0.48\textwidth}
\centering
\begin{tabular}{cc|cc|cc}
 & & \multicolumn{2}{c|}{$y=0$} & \multicolumn{2}{c}{$y=1$} \\
 & & $b=+1$ & $b=-1$ & $b=+1$ & $b=-1$ \\ \hline
\multirow{2}{*}{$x=0$}
 & $a=+1$ & $1-\zeta$ & \cellcolor{chshcol}$\beta+\gamma+\varepsilon$ & \cellcolor{chshcol}$1-\alpha-\gamma-\delta$ & $0$ \\
 & $a=-1$ & \cellcolor{chshcol}$\alpha+\gamma+\delta$ & $0$ & $\alpha+\gamma+\delta$ & \cellcolor{chshcol}$0$ \\ \hline
\multirow{2}{*}{$x=1$}
 & $a=+1$ & \cellcolor{chshcol}$1-\beta-\gamma-\varepsilon$ & $\varepsilon$ & \cellcolor{chshcol}$1-\beta-\gamma$ & $0$ \\
 & $a=-1$ & $0$ & \cellcolor{chshcol}$\beta+\gamma$ & $\beta+\gamma$ & \cellcolor{chshcol}$0$ \\
\end{tabular}
\caption{Strategy (I,II).}
\end{minipage}
\end{table}

\begin{table}[h!]
\centering
\begin{minipage}{0.48\textwidth}
\centering
\begin{tabular}{cc|cc|cc}
 & & \multicolumn{2}{c|}{$y=0$} & \multicolumn{2}{c}{$y=1$} \\
 & & $b=+1$ & $b=-1$ & $b=+1$ & $b=-1$ \\ \hline
\multirow{2}{*}{$x=0$}
 & $a=+1$ & $1-\alpha-\gamma-\delta$ & \cellcolor{chshcol}$0$ & \cellcolor{chshcol}$1-\alpha-\gamma-\delta$ & $0$ \\
 & $a=-1$ & \cellcolor{chshcol}$\alpha+\gamma+\delta$ & $0$ & $\delta$ & \cellcolor{chshcol}$\alpha+\gamma$ \\ \hline
\multirow{2}{*}{$x=1$}
 & $a=+1$ & \cellcolor{chshcol}$1-\beta-\gamma$ & $0$ & \cellcolor{chshcol}$1-\alpha-\beta-\gamma$ & $\alpha$ \\
 & $a=-1$ & $\beta+\gamma$ & \cellcolor{chshcol}$0$ & $\beta$ & \cellcolor{chshcol}$\gamma$ \\
\end{tabular}
\caption{Strategy (I,III).}
\end{minipage}
\hfill
\begin{minipage}{0.48\textwidth}
\centering
\begin{tabular}{cc|cc|cc}
 & & \multicolumn{2}{c|}{$y=0$} & \multicolumn{2}{c}{$y=1$} \\
 & & $b=+1$ & $b=-1$ & $b=+1$ & $b=-1$ \\ \hline
\multirow{2}{*}{$x=0$}
 & $a=+1$ & $1-\zeta$ & \cellcolor{chshcol}$\beta+\gamma+\varepsilon$ & \cellcolor{chshcol}$1-\alpha-\gamma-\delta$ & $0$ \\
 & $a=-1$ & \cellcolor{chshcol}$\alpha+\gamma+\delta$ & $0$ & $\delta$ & \cellcolor{chshcol}$\alpha+\gamma$ \\ \hline
\multirow{2}{*}{$x=1$}
 & $a=+1$ & \cellcolor{chshcol}$1-\beta-\gamma-\varepsilon$ & $\beta+\gamma+\varepsilon$ & \cellcolor{chshcol}$1-\alpha-\gamma$ & $\alpha+\gamma$ \\
 & $a=-1$ & $0$ & \cellcolor{chshcol}$0$ & $0$ & \cellcolor{chshcol}$0$ \\
\end{tabular}
\caption{Strategy (II,I).}
\end{minipage}
\end{table}

\begin{table}[h!]
\centering
\begin{minipage}{0.48\textwidth}
\centering
\begin{tabular}{cc|cc|cc}
 & & \multicolumn{2}{c|}{$y=0$} & \multicolumn{2}{c}{$y=1$} \\
 & & $b=+1$ & $b=-1$ & $b=+1$ & $b=-1$ \\ \hline
\multirow{2}{*}{$x=0$}
 & $a=+1$ & $1-\zeta$ & \cellcolor{chshcol}$\beta+\gamma+\varepsilon$ & \cellcolor{chshcol}$1-\alpha-\gamma-\delta$ & $0$ \\
 & $a=-1$ & \cellcolor{chshcol}$\alpha+\gamma+\delta$ & $0$ & $\alpha+\gamma+\delta$ & \cellcolor{chshcol}$0$ \\ \hline
\multirow{2}{*}{$x=1$}
 & $a=+1$ & \cellcolor{chshcol}$1-\beta-\gamma-\varepsilon$ & $\beta+\gamma+\varepsilon$ & \cellcolor{chshcol}$1$ & $0$ \\
 & $a=-1$ & $0$ & \cellcolor{chshcol}$0$ & $0$ & \cellcolor{chshcol}$0$ \\
\end{tabular}
\caption{Strategy (II,II).}
\end{minipage}
\hfill
\begin{minipage}{0.48\textwidth}
\centering
\begin{tabular}{cc|cc|cc}
 & & \multicolumn{2}{c|}{$y=0$} & \multicolumn{2}{c}{$y=1$} \\
 & & $b=+1$ & $b=-1$ & $b=+1$ & $b=-1$ \\ \hline
\multirow{2}{*}{$x=0$}
 & $a=+1$ & $1-\alpha-\gamma-\delta$ & \cellcolor{chshcol}$0$ & \cellcolor{chshcol}$1-\alpha-\gamma-\delta$ & $0$ \\
 & $a=-1$ & \cellcolor{chshcol}$\alpha+\gamma+\delta$ & $0$ & $\delta$ & \cellcolor{chshcol}$\alpha+\gamma$ \\ \hline
\multirow{2}{*}{$x=1$}
 & $a=+1$ & \cellcolor{chshcol}$1$ & $0$ & \cellcolor{chshcol}$1-\alpha-\gamma$ & $\alpha+\gamma$ \\
 & $a=-1$ & $0$ & \cellcolor{chshcol}$0$ & $0$ & \cellcolor{chshcol}$0$ \\
\end{tabular}
\caption{Strategy (II,III).}
\end{minipage}
\end{table}

\begin{table}[h!]
\centering
\begin{minipage}{0.48\textwidth}
\centering
\begin{tabular}{cc|cc|cc}
 & & \multicolumn{2}{c|}{$y=0$} & \multicolumn{2}{c}{$y=1$} \\
 & & $b=+1$ & $b=-1$ & $b=+1$ & $b=-1$ \\ \hline
\multirow{2}{*}{$x=0$}
 & $a=+1$ & $1-\beta-\gamma-\varepsilon$ & \cellcolor{chshcol}$\beta+\gamma+\varepsilon$ & \cellcolor{chshcol}$1-\alpha-\gamma$ & $\alpha+\gamma$ \\
 & $a=-1$ & \cellcolor{chshcol}$0$ & $0$ & $0$ & \cellcolor{chshcol}$0$ \\ \hline
\multirow{2}{*}{$x=1$}
 & $a=+1$ & \cellcolor{chshcol}$1-\beta-\gamma-\varepsilon$ & $\varepsilon$ & \cellcolor{chshcol}$1-\alpha-\beta-\gamma$ & $\alpha$ \\
 & $a=-1$ & $0$ & \cellcolor{chshcol}$\beta+\gamma$ & $\beta$ & \cellcolor{chshcol}$\gamma$ \\
\end{tabular}
\caption{Strategy (III,I).}
\end{minipage}
\hfill
\begin{minipage}{0.48\textwidth}
\centering
\begin{tabular}{cc|cc|cc}
 & & \multicolumn{2}{c|}{$y=0$} & \multicolumn{2}{c}{$y=1$} \\
 & & $b=+1$ & $b=-1$ & $b=+1$ & $b=-1$ \\ \hline
\multirow{2}{*}{$x=0$}
 & $a=+1$ & $1-\beta-\gamma-\varepsilon$ & \cellcolor{chshcol}$\beta+\gamma+\varepsilon$ & \cellcolor{chshcol}$1$ & $0$ \\
 & $a=-1$ & \cellcolor{chshcol}$0$ & $0$ & $0$ & \cellcolor{chshcol}$0$ \\ \hline
\multirow{2}{*}{$x=1$}
 & $a=+1$ & \cellcolor{chshcol}$1-\beta-\gamma-\varepsilon$ & $\varepsilon$ & \cellcolor{chshcol}$1-\beta-\gamma$ & $0$ \\
 & $a=-1$ & $0$ & \cellcolor{chshcol}$\beta+\gamma$ & $\beta+\gamma$ & \cellcolor{chshcol}$0$ \\
\end{tabular}
\caption{Strategy (III,II).}
\end{minipage}
\end{table}

\begin{table}[h!]
\centering
\small
\begin{tabular}{cc|cc|cc}
 & & \multicolumn{2}{c|}{$y=0$} & \multicolumn{2}{c}{$y=1$} \\
 & & $b=+1$ & $b=-1$ & $b=+1$ & $b=-1$ \\ \hline
\multirow{2}{*}{$x=0$}
 & $a=+1$ & $1$ & \cellcolor{chshcol}$0$ & \cellcolor{chshcol}$1-\alpha-\gamma$ & $\alpha+\gamma$ \\
 & $a=-1$ & \cellcolor{chshcol}$0$ & $0$ & $0$ & \cellcolor{chshcol}$0$ \\ \hline
\multirow{2}{*}{$x=1$}
 & $a=+1$ & \cellcolor{chshcol}$1-\beta-\gamma$ & $0$ & \cellcolor{chshcol}$1-\alpha-\beta-\gamma$ & $\alpha$ \\
 & $a=-1$ & $\beta+\gamma$ & \cellcolor{chshcol}$0$ & $\beta$ & \cellcolor{chshcol}$\gamma$ \\
\end{tabular}
\caption{Strategy (III,III).}
\end{table}

\section{Proofs}\label{app:proof}

\subsection{Explicit probabilities}\label{app:probabilities}

We report the explicit form of the probabilities defined in Eq.~\eqref{eq:prob_def}. We start with the probability that the state reaches party $k$ without having been measured. The state is not measured if all previous parties received input $1$ and chose not to measure. Thus
\begin{align}\label{eq:qno}\notag
    q_k^{no}&=\frac{1}{2}(1-p_1) \cdot \frac{1}{2}(1-p_2) \cdots \frac{1}{2}(1-p_{k-1})\\
    &= \prod_{i=1}^{k-1}\frac{1}{2}(1-p_i)=\frac{1}{2^{k-1}}\prod_{i=1}^{k-1}(1-p_i).
\end{align}

Then, the probability that the system has already been measured in the second basis (the one for input $1$) before reaching party $k$ is equivalent to the probability that any of the previous $1,\dots,k-1$ parties receives the unmeasured state, has input $1$, and measures it. More concretely, this can happen when party $1$ has input $1$ and measures; or party $1$ has input $1$, does not measure, but party $2$ also has input $1$ and measures; and so on. Hence:
\begin{align}\label{eq:qM1}\notag
    q_k^{M_1} =& \frac{1}{2}p_1 + \frac{1}{4}(1-p_1)p_2 + \frac{1}{8}(1-p_1)(1-p_2)p_3+ \dots + \frac{1}{2^{k-1}}\left(\prod_{i=1}^{k-2}(1-p_i)\right) p_{k-1} \\
    =& \sum_{i=1}^{k-1}\frac{1}{2^i}\left(\prod_{j=1}^{i-1}(1-p_j)\right) p_i.
\end{align}

Finally, the probability that the state was measured in the first basis (input $0$) before position $k$ is obtained from similar considerations. In particular, it happens if the first party has input $0$, or if the first party has input $1$, decides not to measure, and party $2$ has input $0$, and so on:
\begin{align}\label{eq:qM0}\notag
    q_k^{M_0} &= \frac{1}{2} + \frac{1}{4}(1-p_1) + \frac{1}{8}(1-p_1)(1-p_2)+ \dots +\frac{1}{2^{k-1}}\prod_{i=1}^{k-2}(1-p_i) \\
    &= \sum_{i=1}^{k-1}\frac{1}{2^i}\prod_{j=1}^{i-1}(1-p_j).
\end{align}

\subsection{Proof of the CHSH condition}\label{app:chsh_proof}
In the main text we established that
\begin{align}\notag
S_{(j,k)}=& \frac{3}{4} + \frac{\gamma}{2} p_j(\text{I}) p_k(\text{I}) - \frac{\alpha}{2}  p_j(\text{III}) p_k(\text{I}) 
- \frac{\beta}{2}  p_j(\text{I}) p_k(\text{III})  -\frac{\alpha+\beta+\gamma}{2}  p_j(\text{III}) p_k(\text{III}) \\\notag
    =& \frac{3}{4} + \frac{\gamma}{2} q_j^{no} p_j \; q_k^{no} p_k - \frac{\alpha}{2} q_j^{M_1} q_k^{no} p_k
    - \frac{\beta}{2} q_j^{no} p_j  q_k^{M_1} -\frac{\alpha+\beta+\gamma}{2} q_j^{M_1} q_k^{M_1}.
\end{align}
Here, we show that $S_{(j,k)}>3/4$ for all pairs $(j,k)$ whenever
\begin{equation}
    q_i^{no} p_i > 3\frac{\alpha+\beta+\gamma}{\gamma}q_i^{M_1} \quad \text{for } i=j,k\, .
\end{equation}
Setting $X_1:=q_j^{no} p_j$, $Y_1:=q_k^{no} p_k$, $X_2:=q_j^{M_1}$, $Y_2:= q_k^{M_1}$, the condition $S_{(j,k)}>3/4$ is equivalent to
\begin{equation}
    \frac{\gamma}{2} X_1 Y_1 - \frac{\alpha}{2} X_2 Y_1 - \frac{\beta}{2} X_1 Y_2 -\frac{\alpha+\beta+\gamma}{2} X_2 Y_2> 0.
\end{equation}
Since $(\alpha+\beta+\gamma)\geq \alpha,\beta,\gamma$ for any $\alpha, \beta, \gamma\in[0,1]$, this inequality holds if the following inequality holds:
\begin{align}
    \frac{\gamma}{\alpha+\beta+\gamma} > \frac{X_2}{X_1} + \frac{Y_2}{Y_1} + \frac{X_2 Y_2}{X_1 Y_1}.
\end{align}
Assuming $X_1 > 3\frac{\alpha+\beta+\gamma}{\gamma} X_2$ and $Y_1 > 3\frac{\alpha+\beta+\gamma}{\gamma} Y_2$, we obtain
\begin{align}
     \frac{X_2}{X_1} + \frac{Y_2}{Y_1} + \frac{X_2 Y_2}{X_1 Y_1}<\frac{2\gamma}{3(\alpha+\beta+\gamma)}+\left(\frac{\gamma}{3(\alpha+\beta+\gamma)}\right)^2 < \frac{2\gamma}{3(\alpha+\beta+\gamma)}+\frac{\gamma}{3(\alpha+\beta+\gamma)}=\frac{\gamma}{\alpha+\beta+\gamma},
\end{align}
which completes the proof.

\subsection{Choice of probabilities}\label{app:prob_choice}

Consider the probabilities
\begin{equation}
    p_i = \frac{\mu^i}{\sum_{l=i}^n\mu^l}=\frac{1-\mu}{1-\mu^{n-i+1}}.
\end{equation}
We verify that this choice satisfies condition~\eqref{eq:prob_condition} for any $\mu > 2 + \frac{3(\alpha+\beta+\gamma)}{\gamma}$. Note that by the definition of $p_i$ we obtain
\begin{align}
    1-p_i=\frac{\mu (1-\mu^{n-i})}{1-\mu^{n-i+1}} \, .
\end{align}
Furthermore, the following identity is useful:
\begin{align}
    \left(\prod_{l=1}^{i-1}(1-p_l)\right)\cdot p_i=\frac{\mu (1-\mu^{n-1})}{1-\mu^{n}}\cdot \frac{\mu (1-\mu^{n-2})}{1-\mu^{n-1}}\cdot ... \cdot \frac{\mu (1-\mu^{n-i+1})}{1-\mu^{n-i+2}}\cdot \frac{1-\mu}{1-\mu^{n-i+1}}=\frac{\mu^{i-1}(1-\mu)}{(1-\mu^n)} \, .
\end{align}
Using this identity, we can compute now $q_i^{no}\,p_i$ using Eq.~\eqref{eq:qno}:
\begin{align}
        q_i^{no}\, p_i
        &= \prod_{l=1}^{i-1}\frac{1}{2}(1-p_l)\cdot p_i
        =\frac{1}{2^{i-1}}p_i\prod_{l=1}^{i-1}(1-p_l)=\left(\frac{\mu}{2}\right)^{i-1}\frac{1-\mu}{1-\mu^n}.
\end{align}
As for $q_i^{M_1}$, we use Eq.~\eqref{eq:qM1} and obtain
\begin{align}
        q_i^{M_1}
        &= \sum_{l=1}^{i-1}\frac{1}{2^l}\prod_{m=1}^{l-1}(1-p_m)p_l
        = \sum_{l=1}^{i-1}\frac{1}{2^l} \frac{\mu^{l-1}(1-\mu)}{1-\mu^n}
        = \frac{1-\mu}{2(1-\mu^n)}\sum_{l=1}^{i-1}\left(\frac{\mu}{2}\right)^{l-1}=\frac{1-\mu}{2(1-\mu^n)}\frac{1-\left(\frac{\mu}{2}\right)^{i-1}}{1-\left(\frac{\mu}{2}\right)}.
\end{align}
Finally, we compute the ratio
\begin{align}
    \frac{q_i^{no}\, p_i}{q_i^{M_1}}
    &=\frac{2(1-\mu^n)}{1-\mu}\frac{1-\left(\frac{\mu}{2}\right)}{1-\left(\frac{\mu}{2}\right)^{i-1}}\cdot\left(\frac{\mu}{2}\right)^{i-1}\frac{1-\mu}{1-\mu^n}
    =\frac{(2-\mu)\left(\frac{\mu}{2}\right)^{i-1}}{1-\left(\frac{\mu}{2}\right)^{i-1}}=\frac{\mu-2}{1-\left(\frac{2}{\mu}\right)^{i-1}}>\mu-2>\frac{3(\alpha+\beta+\gamma)}{\gamma},
\end{align}
for any $\mu>2+\frac{3(\alpha+\beta+\gamma)}{\gamma}$ and $i>1$, as required.

\section{Equivalence between qutrit and qubit protocols}\label{app:bit_equivalence}
In this appendix, we show that the qutrit ancilla used in the main text can be replaced by a qubit ancilla supplemented by shared classical randomness, while preserving the resulting correlations. The shared random variable $\lambda\in\{1,\ldots,n\}$ determines which observer uses strategy~(I); all others use strategy~(II) or~(III) depending solely on the state of the incoming ancilla. Focusing on Alice's side: Alice-1 applies strategy~(I) if $\lambda=1$ and strategy~(II) otherwise. The outgoing qubit ancilla is then prepared in the state $\ket{x_1}$. For $i>1$, Alice-$i$ proceeds as follows:
\begin{itemize}
\item If the incoming ancilla is $\ket0$: apply strategy~(II) and prepare the outgoing ancilla in $\ket0$.
\item If the incoming ancilla is $\ket1$ and $\lambda>i$: apply strategy~(II) and prepare the outgoing ancilla in $\ket{x_i}$.
\item If the incoming ancilla is $\ket1$ and $\lambda=i$: apply strategy~(I) and prepare the outgoing ancilla in $\ket{x_i}$.
\item If the incoming ancilla is $\ket1$ and $\lambda<i$: apply strategy~(III) and prepare the outgoing ancilla in $\ket1$.
\end{itemize}
Thus, the ancilla state $\ket0$ signals the system was measured with $M_0$, while the state $\ket1$ indicates either that the subsystem has not yet been measured (if $\lambda\geq i$) or has been measured with $M_1$ (if $\lambda<i$). Together with the shared random variable $\lambda$ this reproduces exactly the information carried by the qutrit ancilla.

We define the distribution of $\lambda$ by
\begin{align}
    p(\lambda \geq k)=\prod_{i=1}^{k-1} (1-p_i),
\end{align}
from which it follows that
\begin{align}
    p(\lambda = k)=p(\lambda \geq k)-p(\lambda \geq k+1)=p_k\prod_{i=1}^{k-1}(1-p_i).
\end{align}

Party $k$ receives an unmeasured system if and only if $\lambda\geq k$ and all previous parties received input $1$:
\begin{align}
    q_k^{no}=\frac{1}{2^{k-1}}\,p(\lambda \geq k)=\frac{1}{2^{k-1}}\prod_{i=1}^{k-1}(1-p_i),
\end{align}
which coincides with Eq.~\eqref{eq:qno}.

Party $k$ receives a system measured with $M_0$ when some party $i<k$ receives input $0$, while all earlier parties receive input $1$ and $\lambda\geq i$:
\begin{align}
    q_k^{M_0}=\sum_{i=1}^{k-1}\frac{1}{2^i}\,p(\lambda \geq i)
    =\sum_{i=1}^{k-1}\frac{1}{2^i}\prod_{j=1}^{i-1}(1-p_j),
\end{align}
in agreement with Eq.~\eqref{eq:qM0}.

Finally, the state has been measured with $M_1$ before reaching party $k$ if all parties up to $i$ receive input $1$ and $\lambda=i$:
\begin{align}
    q_k^{M_1}=\sum_{i=1}^{k-1}\frac{1}{2^i}\,p(\lambda=i)
    =\sum_{i=1}^{k-1}\frac{1}{2^i}\prod_{j=1}^{i-1}(1-p_j)p_i,
\end{align}
which matches Eq.~\eqref{eq:qM1}. Therefore, the probabilities in the qubit protocol coincide with those of the qutrit protocol, and both implementations lead to the same correlations.

If the probabilities $p_i$ are chosen as in Eq.~\eqref{eq:prob}, the distribution of $\lambda$ becomes
\begin{equation}
    p(\lambda=i)=\frac{\mu^i}{\sum_{l=1}^{n}\mu^l},
\end{equation}
where $\mu>2+\frac{3(\alpha+\beta+\gamma)}{\gamma}$.

\end{document}